\documentclass[aps,prb,twocolumn,superscriptaddress,floatfix,preprintnumbers,showpacs]{revtex4-1}
\usepackage{graphicx,graphics,amssymb,amsmath,epsfig,color,subcaption,mwe,soul,lipsum}
\usepackage[export]{adjustbox}
\usepackage{inputenc}
\DeclareMathOperator{\tr}{tr}
\DeclareMathOperator{\noise}{N}
\DeclareMathOperator{\susc}{{\rm a}}
\DeclareMathOperator{\noiseb}{ N}
\DeclareMathOperator{\snoise}{\overline{ N}}

\DeclareMathOperator{\spin}{\hat{S}}
\DeclareMathOperator{\spinb}{\hat{\bold S}}
\DeclareMathOperator{\hO}{\hat{O}}

\begin{document}

\title{Interaction without back-action in the context of quantum manipulation}
\author{V.P. Michal}
\affiliation{QuTech and Kavli Institute of Nanoscience, TU Delft, 2600 GA Delft, The Netherlands}
\author{Y.V. Nazarov}
\affiliation{Kavli Institute of Nanoscience, Delft University of Technology,
Lorentzweg 1, NL-2628 CJ, Delft, The Netherlands}

\begin{abstract}
We address the interaction between two quantum systems (A and B) that is mediated by their common linear environment. If the environment is out of equilibrium the resulting interaction violates Onsager relations and cannot be described by a Hamiltonian. In simple terms the action of system A on system B does not necessarily produce a back-action. We derive general quantum equations describing the situation and analyze in details their classical correspondence. Changing the properties of the environment one can easily change and engineer the resulting interaction. It is tempting to use this for quantum manipulation of the systems. However the resulting quantum gate is not always unitary and may induce a loss of quantum coherence. For a relevant example we consider systems A and B to be spins of arbitrary values and arrange the interaction to realize an analogue of the two-qubit CNOT gate. The direction of spin A controls the rotation of spin B while spin A is not rotated experiencing no back-action from spin B. We solve the quantum dynamics equations and analyze the purity of the resulting density matrix. The resulting purity essentially depends on the initial states of the systems. We attempt to find a universal characteristics of the purity optimizing it for the worst choice of initial states. For both spins $s_A=s_B=1/2$, the optimized purity is bounded by 1/2 irrespective of the details of the gate. We also study in detail the semiclassical limit of large spins. In this case the optimized purity is bounded by $(1+\pi/2)^{-1}\approx0.39$. This is much better than the typical purity of a large spin state $\sim s^{-1}$. We conclude that although the quantum manipulation without back-action inevitably causes decoherence of the quantum states the actual purity of the resulting state can be optimized and made relatively high.
\end{abstract}

\maketitle

\section{Introduction}

The ability to control open quantum systems is important in the context of information processing, that includes information processing with spins \cite{Wolf2001,Awschalom2002,Zutic2004} and spin quantum computation \cite{QuantumInformation,LDV,Kane,HansonRMP,ZwanenburgRMP}. In general the environment is harmful to quantum systems as it leads to dissipation and the loss of the quantum coherence. On the other hand coupling the system to an environment is essential in the quantum manipulation itself\cite{Moiseyev,Uzdin2013,Muller2012,Emary2013,Bender2010,Evans2015,Ketterer2014}. 

Coupling quantum systems together can be achieved in many ways depending on the experimental platform \cite{Petta2005,Niskanen2007,Nowack2011,Braakman2013,Srinivasa2015,Baart2016}.
The environment can mediate the interaction between the subsystems as described by linear response theory \cite{Kubo}. If the environment is in thermal equilibrium the environment-mediated interaction between two systems A and B is governed by the Onsager symmetries\cite{Onsager,VanKampen}: in simple terms the action that system A has on system B is equivalent to the action of system B on system A. Such a dissipative interaction was studied in Refs. \citenum{Benatti2003,Lai2008,Sun2018}.

Here we examine the more general case where the environment is out of thermal equilibrium and the Onsager symmetries do not hold. In particular this makes possible the situation where subsystem A influences the dynamics of system B and the reciprocal interaction does not occur. We call this: interaction without back-action. By solving the equations of the dynamics of spins in an environment we show that interaction without back-action inevitably leads to noise and decoherence of the subsystems. We also investigate if this interaction can realize the controlled-NOT (C-NOT gate) that is central to quantum information processing\cite{Monroe1995,Veldhorst2015,Zajac2018}. 
We find that this is possible if the number of levels involved is large (large spin quantum number). In contrast for two-level systems (spin $s=1/2$) the coherence is almost entirely lost. 
The C-NOT operation is defined for a pair of $s=1/2$ spins. Let us note that it can be naturally generalized to the case of bigger spins, $s_A$ and $s_B$. During the operation $\langle\spinb_A\rangle$ remains unchanged while $\langle\spinb_B\rangle$ is rotated about a certain axis by an angle $\pm\pi/2$, the sign being dependent on the direction of $\langle\spinb_A\rangle$. Such a generalization naturally arises if one applies the C-NOT operation to multiple copies of the pair of $s=1/2$ spins with the same initial state in each pair. This is a usual approach in error correction schemes \cite{QuantumInformation}.  

The paper is organized as follows: in the second part we discuss the classical dynamics of two systems in the presence of a non-reciprocal interaction between the systems. In the third part we address the dynamics of classical spins. In the fourth part we derive the equations of dynamics of open quantum systems. In the fifth part we compare the quantum evolution of observables with the classical limit. In the sixth part we compute the time evolution of spin states in a simple model where dephasing is taken into account and we study the feasibility of the C-NOT operation for spin-1/2 and for large spins. The last section is dedicated to the conclusion.

\section{General classical dynamics}

Before addressing the environment-mediated interaction at the quantum level, let us consider it at the classical level.

The description of interaction at this level is very simple and compact. The environment can be regarded as a piece of electronics that functionally consists of a set of controls that exert forces on the system variables, a set of meters that measure the variables, and provides feedback setting the control forces corresponding to the measurement results. We will assume an instant and linear feedback. In this case, the environment and the mediated interaction is completely characterized by a matrix of linear susceptibilities $\susc_{AB}$, 
$\susc_{AB}$ being the proportionality coefficient between the force conjugated to the variable $A$ and the value of the variable $B$. In principle, the environment can be always engineered to provide any desired $\susc_{AB}$.

The symmetry properties of this matrix are important. If $\susc_{AB}=\susc_{BA}$ the resulting dynamics is Hamiltonian, and there is a conserving energy characteristic to the dynamics. The symmetry is equivalent to the Newton's third law: every action produces equal and opposite reaction. 

If the environment is in thermodynamic equilibrium, the symmetry property as well as the energy conservation is guaranteed by Onsager symmetry principle \cite{Onsager}. An active environment is not in the equilibrium, for instance, the piece of electronics realizing  $\susc_{AB}$ may be connected to the battery, and it should lead to $\susc_{AB} \ne \susc_{BA}$.
In this paper, we concentrate on the cases when Onsager relation is violated.

To illustrate the importance of Onsager symmetry let us consider the classical dynamics of two identical oscillators with coordinates $q_{A,B}$ that are weakly coupled via the environment
\begin{subequations}
\begin{align}
&m\ddot{q}_A+m\omega^2q_A+\susc_{AB}q_B=0,\label{classdyn1}\\
&m\ddot{q}_B+m\omega^2q_B+\susc_{BA}q_A=0.\label{classdyn2}
\end{align}
\end{subequations}
$\omega$ being the oscillator frequency, $\susc_{AB,BA} \ll m \omega^2$. 
These linear dynamics are easy to analyze. 

If $\susc_{AB}\susc_{BA} >0$, there are two close oscillating frequencies $\omega \pm (2m\omega)^{-1} \sqrt{\susc_{AB}\susc_{BA}}$. The absence of Onsager symmetry is only manifested in non-symmetric eigenvectors of the corresponding oscillating modes. If $\susc_{AB}\susc_{BA} <0$, the oscillators become unstable, there is an exponentially growing oscillation with amplitude 
$\propto \exp(t (2m\omega)^{-1} \sqrt{-\susc_{AB}\susc_{BA}})$. Apparently, the energy is supplied by the environment. Another interesting case is $\susc_{BA}=0$. In this case, the motion of the oscillator $B$ is not affected by the coupling. The oscillator $A$ experiences the oscillations of $B$ as an external resonance force, so its oscillation amplitude increases linearly with time. One can say that the oscillator $A$ provides an unobtrusive and very efficient detection and amplification of $B$.

This example illustrates the importance of the Onsager symmetry violation at the classical level: it may drastically change the dynamics, but does not have to do this always.

Let us consider the most general classical equations of motion in the linear response regime. 
Let us have a set of classical variables $O_\alpha$ (they may be distibuted between two coupled systems). 
Their evolution is governed by the equation,
\begin{equation}\label{dotOclass}
\dot{O}_\alpha=\{O_\alpha,H\}_{pb}+\susc_{\beta\gamma}\{O_\alpha,O_\beta\}_{pb}O_\gamma.
\end{equation}
The first term includes the Hamiltonian contribution to the system dynamics and $\{...\}_{pb}$ stands for Poisson brackets,
\begin{equation}
\{f,g\}_{pb}=\frac{\partial f}{\partial q_\gamma}\frac{\partial g}{\partial p_\gamma}-\frac{\partial f}{\partial p_\gamma}\frac{\partial g}{\partial q_\gamma}
\end{equation}
$q_\gamma,p_\gamma$ being the canonical coordinates of the system. This is the most general linear feedback equation that will be compared in Section \ref{quantum} with quantum equations in order to establish classical/quantum correspondence.
\
The system of equations (\ref{dotOclass}) is closed if the Poisson brackets of $O_\alpha$ can be expressed in terms of the elements of the set of constants. This is obviously the case when $O_\alpha$ are canonical variables. The Poisson brackets in this case are $\{q_j,q_k\}_{pb}=0$, $\{p_j,p_k\}_{pb}=0$, $\{q_j,p_k\}_{pb}=\delta_{jk}$ ($j,k=A,B$), and the interaction terms are expressed as follows:
\begin{subequations}
\begin{align}
&\dot{q}_j=\susc_{p_jq_k}q_k+\susc_{p_jp_k}p_k,\label{ps1}\\
&\dot{p}_j=-\susc_{q_jq_k}q_k-\susc_{q_jp_k}p_k.\label{ps2}
\end{align}
\end{subequations}
The dynamics given by Eqs. (\ref{classdyn1}), (\ref{classdyn2}) previously discussed is a particular case of Eqs. (\ref{ps1}) and (\ref{ps2}).

While the quantum version of these oscillator equations is worth exploring, its relation to quantum information processing is not direct since the resulting equations are linear. Another example of a variable set where Poisson brackets are closed is provided by classical spins. Since the resulting equations are not linear and there is a direct and straightforward analogy between the spins and (collections of) qubits, we concentrate on this case.
 
\section{Classical spins}

Let us define two classical spins through angular momenta expressed in therms of canonical variables $\mathbf{S}_j=\mathbf{q}_j\times\mathbf{p}_j$ ($j=A,B$). The Poisson brackets read 
\begin{equation}
\{S^a_i,S^b_j\}_{pb} = \epsilon_{abc}\delta_{ij}S_i^c,
\end{equation}
with $\epsilon_{abc}$ the Levi-Civita antisymmetric tensor and $\delta_{jk}$ the Kronecker symbol.
The susceptibility matrix $\susc_{AB}^{ab}$ relates the "magnetic field" acting on spin $A$ to the spin components of spin $B$. The resulting equations of motion read 
\begin{subequations}
\begin{align}
\dot{S}_A^a=\epsilon_{abc}\susc_{AB}^{bd}S_B^dS_A^c\label{dotl1} \\
\dot{S}_B^a=\epsilon_{abc}\susc_{BA}^{bd}S_A^dS_B^c\label{dotl2}
\end{align}
\end{subequations}
These equations conserve the moduli of both spins, $|{\bf S}_{A,B}|^2=\textrm{const}$. 
If Onsager symmetry holds, the quantity $S_A^a \susc_{AB}^{ab} S_B^b$ is conserved as well, and the equations have stable stationary solutions. This is not generally the case if Onsager symmetry is violated.

Let us concentrate on a simple extreme case where $\susc_{BA}^{db}=0$ and $\susc_{AB}^{bd}\neq0$. In this case, the spin $B$ remains constant in time. It creates a constant magnetic field for the spin $A$ that precesses around this magnetic field.

This behavior is reminiscent to that of the C-NOT gate that is of fundamental importance in quantum information processing \cite{QuantumInformation}. This is a two-qubit operation that does not modify the spin of the first (control) qubit. The spin of the second qubit does not change if the first spin is "up" and flips its direction if the first qubit is "down"

Let us note that we can realize a classical analogue of the C-NOT gate by means of the environment-mediated interaction. We assume that we can switch the interaction on and off. We arrange interaction such that the only non-vanishing element of the susceptibility matrix is $\susc_{AB}^{zz} \equiv \hbar\nu$. We switch on the interaction for a time interval $\Delta t$ such that the spin $A$ is rotated by $\pm \pi/2$ about z-axis provided the spin $B$ is in $\pm z$ direction, $\Delta t = \pi/2\nu$. We switch off the interaction and apply "magnetic field" to the spin $A$ to rotate it by $\pi/2$.

Thereby we achieve the controlled-NOT gate functionality. It is worth noting that big spins $A,B$ can be seen as collections of aligned qubit spins. The operation is performed on all qubits of the collection and they remain aligned at each stage of the operation. 

\section{Open system quantum dynamics}\label{quantum}

In this section we derive the Bloch-Master equations that describe the dynamics of our system. Our derivation in main follows the standard lines, see e.g. Ref. (\citenum{Benatti2003}) and references therein. However we need to account for the fact that the environment is not in thermal equilibrium. We also group the terms differently to facilitate the comparison with the classical equations (\ref{dotOclass}).

The ensemble system + bath is described by a density matrix $\rho$ whose time evolution is given by the von Neumann equation
\begin{equation}
\dot{\rho}(t)=-\frac{i}{\hbar}[H,\rho(t)]=-\frac{i}{\hbar}(H\rho(t)-\rho(t)H).
\end{equation}
The Hamiltonian of the system and the bath reads
\begin{equation}
H=H_s+H_b+H_c.
\end{equation}
It includes the Hamiltonian of the subsystem $H_s$, the Hamiltonian of the bath $H_b$, and the Hamiltonian $H_c$ that describes the coupling between the two. The latter is written as
\begin{equation}
H_c=-\hO_\alpha Q_\alpha,
\end{equation}
where $\hO_\alpha$ are the operators of the subsystem, $Q_\alpha$ are the operators of the bath, and there is summation over the repeated index.
The density matrix of the subsystem is obtained by taking the trace over the bath variables of the full density matrix:
\begin{equation}\label{rhos}
\rho_s(t)=\tr_b\{\rho(t)\}.
\end{equation}
Let us define
\begin{equation}\label{rhot}
\tilde{\rho}(t)=e^{i tH_b/\hbar}\rho(t)e^{-i tH_b/\hbar}.
\end{equation}
Assuming $H_b$ to be time-independent we can check that $\tilde{\rho}(t)$ satisfies the equation
\begin{equation}\label{rhotd}
\dot{\tilde{\rho}}(t)=-\frac{i}{\hbar}[H_s(t),\tilde{\rho}(t)]-\frac{i}{\hbar}[\tilde{H}_c(t),\tilde{\rho}(t)],
\end{equation}
where the transformed Hamiltonian $\tilde{H}_{c}$ reads
\begin{eqnarray}\label{transfo}
\nonumber\tilde{H}_{c}(t)&=&e^{i tH_b/\hbar}H_{c}e^{-i tH_b/\hbar}\\
\nonumber&=&-\hO_\alpha e^{i tH_b/\hbar}Q_\alpha e^{-i tH_b/\hbar}\\
\nonumber&=&-\hO_\alpha\tilde{Q}_\alpha(t).
\end{eqnarray}
Integrating Eq.(\ref{rhotd}) leads to
\begin{equation}\label{rhot2}
\tilde{\rho}(t)=\tilde{\rho}(0)-\frac{i}{\hbar}\int_0^tdt'[H_s(t')+\tilde{H}_c(t'),\tilde{\rho}(t')],
\end{equation}
and inserting this into the second term of Eq.(\ref{rhotd}) gives
\begin{eqnarray}
\nonumber\dot{\tilde{\rho}}(t)&=&-\frac{i}{\hbar}[H_s(t),\tilde{\rho}(t)]-\frac{i}{\hbar}[\tilde{H}_c(t),\tilde{\rho}(0)]\\
&&-\frac{1}{\hbar^2}\int_0^tdt'\,[\tilde{H}_c(t),[H_s(t')+\tilde{H}_c(t'),\tilde{\rho}(t')]].
\end{eqnarray}
At this stage we let ${\tilde\rho}(t)\to\tilde{\rho}_b(t)\otimes\rho_s(t)$. Taking the trace over the bath variables yields
\begin{eqnarray}
\nonumber\dot{\rho}_s(t)&=&-\frac{i}{\hbar}[H_s(t),\rho_s(t)]\\
\nonumber&&-\frac{1}{\hbar^2}\int_0^tdt'\tr_b\{\tilde{Q}_\alpha(t)\tilde{Q}_\beta(t')\tilde{\rho}_b(t')\}\hO_\alpha \hO_\beta\rho_s(t')\\
\nonumber&&+\frac{1}{\hbar^2}\int_0^tdt'\tr_b\{\tilde{Q}_\beta(t)\tilde{\rho}_b(t')\tilde{Q}_\alpha(t')\}\hO_\beta\rho_s(t')\hO_\alpha\\
\nonumber&&+\frac{1}{\hbar^2}\int_0^tdt'\tr_b\{\tilde{Q}_\beta(t')\tilde{\rho}_b(t')\tilde{Q}_\alpha(t)\}\hO_\beta\rho_s(t')\hO_\alpha\\
\nonumber&&-\frac{1}{\hbar^2}\int_0^tdt'\tr_b\{\tilde{\rho}_b(t')\tilde{Q}_\alpha(t')\tilde{Q}_\beta(t)\}\rho_s(t')\hO_\alpha \hO_\beta.
\end{eqnarray}
Let $\langle\cdot\rangle_b=\tr_b\{\tilde{\rho}_b(t)\cdot\}$.
The bath operators vanish on average: $\langle\tilde{Q}_\alpha(t)\rangle_b=0$, and the correlators $\langle\tilde{Q}_\alpha(t)\tilde{Q}_\beta(t')\rangle_b$ decay fast for $t-t'>t_b$, $t_b$ being the time scale of the correlations of the bath variables. We also assume $t\gg t_b$, the elements of the density matrix are essentially constant over the time scale $t_b$, and the correlators are uniform in time. Therefore we let
\begin{align}
\nonumber&\int_0^t dt'\langle\tilde{Q}_\alpha(t)\tilde{Q}_\beta(t')\rangle_b \hO_\alpha \hO_\beta\rho_s(t')\\
\nonumber&\to\int_{-\infty}^0dt'\langle\tilde{Q}_\alpha(0)\tilde{Q}_\beta(t')\rangle_b \hO_\alpha \hO_\beta\rho_s(t),\text{ etc.}
\end{align}
Hence the separation of time scales justifies the Born-Markov approximation. By defining
\begin{equation}
\noise_{\alpha\beta}^{-}=\int_{-\infty}^0dt\,\langle\tilde{Q}_\alpha(0)\tilde{Q}_\beta(t)\rangle_b,
\end{equation}
and
\begin{equation}
\noise_{\alpha\beta}^{+}=\int_{-\infty}^0dt\,\langle\tilde{Q}_\alpha(t)\tilde{Q}_\beta(0)\rangle_b,
\end{equation}
we arrive at the differential equation that describes the time evolution of the subsystem density matrix:
\begin{align}
\nonumber\dot{\rho}_s(t)=&-\frac{i}{\hbar}[H_s(t),\rho_s(t)]\\
\nonumber&-\frac{1}{\hbar^2}\noise_{\alpha\beta}^-\hO_\alpha \hO_\beta\rho_s(t)-\frac{1}{\hbar^2}\noise_{\alpha\beta}^+\rho_s(t)\hO_\alpha \hO_\beta\\
&+\frac{1}{\hbar^2}(\noise_{\alpha\beta}^-+\noise_{\alpha\beta}^+)\hO_\beta\rho_s(t)\hO_\alpha.\label{tevol}
\end{align}

Let us introduce the (symmetrized) noises and the susceptibilities
\begin{subequations}
\begin{align}
\label{noise}&\noise_{\alpha\beta}=\noise_{\alpha\beta}^-+\noise_{\beta\alpha}^+\\
\label{snoise}&\snoise_{\alpha\beta}=\frac{1}{2}(\noise_{\alpha\beta}+\noise_{\beta\alpha}),\\
\label{susc}&\susc_{\alpha\beta}=-\frac{i}{\hbar}(\noise_{\alpha\beta}^{-}-\noise_{\beta\alpha}^{+}).
\end{align}
\end{subequations}
Those representations were already presented in literature, see for instance Ref. \citenum{Safi}.
Then we write Eq.(\ref{tevol}) in the form of the standard Bloch master equations:
\begin{eqnarray}\label{lind}
\nonumber\dot{\rho}_s(t)&=&-\frac{i}{\hbar}[H_s'(t),\rho_s(t)]\\
&&-\frac{C_{\alpha\beta}}{2\hbar^2}\big(\{\hO_\alpha \hO_\beta,\rho_s(t)\}-2\hO_\beta\rho_s(t)\hO_\alpha\big)
\end{eqnarray}
where $\{a,b\}=ab+ba$ and
\begin{eqnarray}
H_s'(t)=H_s(t)-\frac{i}{2\hbar}(\noise_{\alpha\beta}^--\noise_{\alpha\beta}^+)\hO_\alpha \hO_\beta,\label{Hsp}\\
C_{\alpha\beta}=\noise_{\alpha\beta}^-+\noise_{\alpha\beta}^+=\snoise_{\alpha\beta}+\frac{i\hbar}{2}(\susc_{\alpha\beta}-\susc_{\beta\alpha}).
\end{eqnarray} 
The matrix $C$ is positive-Hermitian. It includes a symmetric part that consists of the noises $\snoise_{\alpha\beta}$ and an anti-symmetric part that includes the elements of the susceptibility matrix.
In order to reveal its physical content we rewrite Eq.(\ref{lind}) as 
\begin{eqnarray}\label{lind2}
\nonumber\dot{\rho}_s(t)&=&-\frac{i}{\hbar}[H_s''(t),\rho_s(t)]\\
\nonumber&&-\frac{\snoise_{\alpha\beta}}{2\hbar^2}[\hO_\alpha,[\hO_\beta,\rho_s(t)]]\\
&&-\frac{i}{2\hbar}\susc_{\alpha\beta}[\hO_\alpha, \{\hO_\beta,\rho_s(t)\}],
\end{eqnarray}
with
\begin{equation}
H_s''(t)=H_s(t)-\frac{i}{4\hbar}(\noise_{\alpha\beta}-\noise_{\beta\alpha})\hO_\alpha \hO_\beta.
\end{equation}

In the following sections we concentrate on spins. 
The spin operators $\spin_j^a$ ($a=x,y,z$) are defined by the commutation relations
\begin{equation}\label{commut}
[\spin_j^a,\spin_k^b]=i\hbar\epsilon_{abc}\delta_{jk}\spin_j^c.
\end{equation} 
Then Eq.(\ref{lind2}) becomes
\begin{eqnarray}\label{lindspin}
\nonumber\dot{\rho}_s(t)&=&-\frac{i}{\hbar}[H''_s(t),\rho_s(t)]\\
\nonumber&&-\frac{\snoise_{jk}^{ab}}{2\hbar^2}[\spin_j^a,[\spin_k^b,\rho_s(t)]]\\
&&-\frac{i}{2\hbar}\susc_{jk}^{ab}[\spin_j^a,\{\spin_k^b,\rho_s(t)\}]
\end{eqnarray}
with
\begin{equation}
H''_s=H_s-\frac{i}{4\hbar}(\noise_{jk}^{ab}-\noise_{kj}^{ba})\spin_j^a\spin_k^b.
\end{equation}
In Sec.(\ref{state}) we will solve the equations for a pair of spins in the simple case where dephasing only is taken into account.

\section{Comparison between classical and quantum dynamics}\label{comparison}

In order to give a quasiclassical interpretation of the equations derived in the preceding section let us compute the time evolution of the expectation value of the operator $\hO_\alpha$. Using the cyclic property of the trace we obtain from Eq.(\ref{lind2}):
\begin{eqnarray}\label{dotO}
\nonumber\langle\dot{\hO}_\alpha\rangle&=&-\frac{i}{\hbar}\langle[\hO_\alpha,H_s''(t)]\rangle-\frac{\snoise_{\beta\gamma}}{2\hbar^2}\langle[[\hO_\alpha,\hO_\beta],\hO_\gamma]\rangle\\
&&-\frac{i}{2\hbar}\susc_{\beta\gamma}\langle\{[\hO_\alpha,\hO_\beta],\hO_\gamma\}\rangle,
\end{eqnarray}
with $\langle\cdot\rangle=\tr_s\{\rho_s(t)\cdot\}$.

The connection between quantum and classical dynamics is generally obtained by employing the correspondence between the commutators and the Poisson brackets:
\begin{equation} 
-\frac{i}{\hbar}[\hO_\alpha,\hO_\beta]\leftrightarrow\{O_\alpha,O_\beta\}_{pb}.
\end{equation}
In the classical case in addition the noise matrix is symmetric: $\noise_{\alpha\beta}=\noise_{\beta\alpha}$. With those prescriptions the time-evolution of the classical variables $O_\alpha$ writes:
\begin{eqnarray}
\dot{O}_\alpha&=&\{O_\alpha,H_s\}_{pb}\nonumber\\
&+&\frac{\noise_{\beta\gamma}}{2}\{\{O_\alpha,O_\beta\}_{pb},O_\gamma\}_{pb}\nonumber\\
&+&\susc_{\beta\gamma}\{O_\alpha,O_\beta\}_{pb}O_\gamma.
\end{eqnarray}
The first and third lines are consistent with Eq.(\ref{dotOclass}) and there is an additional noise term given by the second line. 

We would like to consider two special cases: canonical operators and spins. 
On the one hand the phase space dynamics is obtained from the commutation relations of the canonical operators   
$[\hat{q}_j^a,\hat{q}_k^b]=0$, $[\hat{p}_j^a,\hat{p}_k^b]=0$, and $[\hat{q}_j^a,\hat{p}_k^b]=i\hbar\delta_{jk}\delta_{ab}=-[\hat{p}_j^a,\hat{q}_k^b]$. The corresponding equations of motion are
\begin{align}
&\langle\dot{\hat{q}}_j^a\rangle=\susc_{p_j^aq_k^b}\langle\hat{q}_k^b\rangle+\susc_{p_j^ap_k^b}\langle\hat{p}_k^b\rangle\\
&\langle\dot{\hat{p}}_j^a\rangle=-\susc_{q_j^aq_k^b}\langle\hat{q}_k^b\rangle-\susc_{q_j^ap_k^b}\langle\hat{p}_k^b\rangle.
\end{align}
They are the quantum counterparts to the classical equations Eqs.(\ref{ps1}) and (\ref{ps2}).

For spins on the other hand, using the commutation relations (\ref{commut}) in Eq.(\ref{dotO}) leads to:
\begin{eqnarray}
\nonumber\langle\dot{\spin}_j^a\rangle&=&-\frac{i}{\hbar}\langle[\spin_j^a,H_s(t)]\rangle+\noiseb_{jj}^{ba}\langle\spin_j^b\rangle-\noiseb_{jj}^{bb}\langle\spin_j^a\rangle\\
&&+\frac{1}{2}\epsilon_{abc}\susc_{jk}^{bd}\langle\{\spin_k^d,\spin_j^c\}\rangle.
\end{eqnarray}
The term $k=j$ can be taken out of the summation, leading to
\begin{eqnarray}\label{dotS}
\nonumber\langle\dot{\spin}_j^a\rangle&=&-\frac{i}{\hbar}\langle[\spin_j^a,H_s(t)]\rangle\\
\nonumber&&+\frac{1}{2}\noiseb_{jj}^{ba}\langle\spin_j^b\rangle-\frac{1}{2}\noiseb_{jj}^{bb}\langle\spin_j^a\rangle\\
\nonumber&&+\frac{1}{2}\epsilon_{abc}\susc_{jj}^{bd}\langle\{\spin_j^d,\spin_j^c\}\rangle\\
&&+\sum_{k\neq j}\epsilon_{abc}\susc_{jk}^{bd}\langle\spin_k^d\spin_j^c\rangle.
\end{eqnarray}
This corresponds to Eq.(9) of Ref.\citenum{Danon2011}.

Spin-1/2 operators in particular satisfy the anticommutation relations $\{\spin_j^d,\spin_j^c\}=\hbar^2\delta_{cd}/2$. In that case the term in the third line is the number $\hbar^2\epsilon_{abc}\susc_{jj}^{bc}/4$. It accounts for the effect of spin pumping\cite{Danon2011}.
The last term of the above equation is the quantum counterpart to Eqs.(\ref{dotl1},\ref{dotl2}).

In Eq.(\ref{dotS}) the noises $\noise_{jj}^{ba}$ lead to relaxation and dephasing of the spins. This is an important feature since in quantum physics the asymmetry of the susceptibility matrix leads to finite noise as required by the positivity of the matrix $C$ in Eq.(\ref{lind}). The noise is expected to reduce the purity of the quantum state. In the next section we compute the purity of the spin quantum states and we find its maximal achievable value in the process of interaction without back-action.

\section{Decoherence during the C-NOT operation: evaluation of purity}\label{state}

In this section we specify a simple model of the environment-induced interaction, describe the C-NOT operation, and evaluate decoherence induced in the course of the operation.
The noises (\ref{noise}) and susceptibilities (\ref{susc}) are:
\begin{subequations}
\begin{align}
&\noise_{jk}^{ab}=\noise_{jk}\delta_{az}\delta_{bz},\\
&\susc_{jk}^{ab}=\susc_{jk}\delta_{az}\delta_{bz}.
\end{align}
\end{subequations}
We consider the two quantum objects (spins) $j=A,B$.

For realizing the C-NOT operation we would like spin A to control the rotation of spin B while spin A remains immobile. This is fulfilled when $\susc_{AB}=0$ and $\susc_{BA}\neq0$.  
If spin A is oriented parallel to the $\hat{z}$ axis then the classical equations of motion for spin B are 
\begin{subequations}
\begin{align}
&\dot{S}_B^x=-\susc_{BA}S_A^z S_B^y,\\
&\dot{S}_B^y=\susc_{BA}S_A^z S_B^x,\\
&\dot{S}_B^z=0.
\end{align}
\end{subequations}
Depending on the orientation of spin A (parallel or anti-parallel to $\hat{z}$), spin B rotates with angular frequency $\pm\susc_{BA}s_A\hat{z}$, where $s_a$... so it rotates by an angle $\pm\pi/2$ over the time 
\begin{equation}\label{time}
t_{\pi/2}=\frac{\pi}{2\susc_{BA}s_A}.
\end{equation}

We would like to evaluate how the quantum states are affected by the noise after time (\ref{time}). A simple measure of the effect of the noise is given by the purity of the quantum state 
\begin{equation}\label{purity}
\gamma(t)=\tr\{\rho_s(t)^2\}.
\end{equation}

To compute the purity we need to evaluate the density matrix at the final time. 
The time-dependence of the density matrix is obtained from Eq.(\ref{lindspin}):  
\begin{eqnarray}
\nonumber\dot{\rho}_s(t)&=&-\frac{\noise_{AA}}{2\hbar^2}[\spin_A^z,[\spin_A^z,\rho_s(t)]]\\
\nonumber&&-\frac{\snoise_{AB}}{2\hbar^2}\big([\spin_A^z,[\spin_B^z,\rho_s(t)]]+[\spin_B^z,[\spin_A^z,\rho_s(t)]]\big)\\
\nonumber&&-\frac{\noise_{BB}}{2\hbar^2}[\spin_B^z,[\spin_B^z,\rho_s(t)]]\\
&&-\frac{i}{2\hbar}\susc_{BA}[\spin_B^z,\{\spin_A^z,\rho_s(t)\}]
\end{eqnarray}
We choose as a basis the products of the eigenstates of $\spin_A^z$ and $\spin_B^z$: $|s_A,m_A\rangle|s_B,m_B\rangle$ with
\begin{eqnarray}\label{basis}
\spin_j^z|s_j,m_j\rangle&=&\hbar m_j|s_j,m_j\rangle,\\
\nonumber m_j&=&-s_j,-s_j+1,\dots,s_j,\\
\nonumber j&=&A,B.
\end{eqnarray}
In this basis we obtain
\begin{equation}\label{dotrhomn}
\dot{\rho}_{\bf mn}(t)=-A_{\bf mn}\rho_{\bf mn}(t),
\end{equation}
with ${\bf m}=(m_A,m_B)$ and
\begin{eqnarray}\label{amn}
\nonumber A_{\bf mn}&=&\frac{1}{2}\noise_{AA}(m_A-n_A)^2+\frac{1}{2}\noise_{BB}(m_B-n_B)^2\\
\nonumber&+&\snoise_{AB}(m_A-n_A)(m_B-n_B)\\
&+&\frac{i\hbar}{2}\susc_{BA}(m_B-n_B)(m_A+n_A).
\end{eqnarray}
These equations show that the noises lead to the decay of the non-diagonal elements of the density matrix which is a property of dephasing. 
The solution to Eq.(\ref{dotrhomn}) is
\begin{equation}\label{rhots}
\rho_{\bf mn}(t)=\rho_{\bf mn}(0)\exp(-tA_{\bf mn}).
\end{equation}
The initial state is taken as the product of the density matrices of pure spin states:
\begin{eqnarray}
\nonumber\rho_{\bf mn}(0)&=&\rho_{m_An_A}(0)\rho_{m_Bn_B}(0)\\
&=&\psi_{m_A}\psi_{n_A}^*\psi_{m_B}\psi_{n_B}^*,
\end{eqnarray}
and we assume the spin coherent states\cite{Lieb73}: 
\begin{equation}
\psi_{m_j}=\Big(C_{s_j+m_j}^{2s_j}p_j^{s_j+m_j}(1-p_j)^{s_j-m_j}\Big)^{1/2}e^{i(s_j-m_j)\phi_j}.
\end{equation}
Here $C_k^n=n!/k!(n-k)!$ are the binomial coefficients and $p_j=(1+\cos\theta_j)/2$. 
These correspond to the states with maximal single-spin projection $s_j$ along the direction $(\theta_j,\phi_j)$, $\theta_j$ and $\phi_j$ being the angles that parametrize the sphere\cite{Lieb73}.
In particular the continuous limit is achieved when $s_j\to\infty$. Applying the central limit theorem to the binomial distribution leads to:
\begin{eqnarray}
\nonumber\psi_{m_j}&\to&\frac{1}{(2\pi s_jq_j)^{1/4}}e^{-\frac{1}{4s_jq_j}(m_j+s_j-2s_jp_j)^2}e^{i(s_j-m_j)\phi_j}\\
&\text{as}&s_j\to\infty,
\end{eqnarray}
with $q_j=2p_j(1-p_j)=(\sin\theta_j)^2/2$. 

\begin{widetext}
The purity Eq. (\ref{purity}) evaluates as
\begin{eqnarray}
\nonumber\gamma(t)&=&\sum_{m_A,n_A=-s_A}^{s_A}\sum_{m_B,n_B=-s_B}^{s_B}|\psi_{m_A}|^2|\psi_{n_A}|^2|\psi_{m_B}|^2|\psi_{n_B}|^2\\
\nonumber&&\times\exp\Big(-t\big(\noise_{AA}(m_A-n_A)^2+2\snoise_{AB}(m_A-n_A)(m_B-n_B)+\noise_{BB}(m_B-n_B)^2\big)\Big),\label{purity2}
\end{eqnarray}
In the large spin regime $s_A=s_B=s\gg1$ it is obtained from the Gaussian integrals:
\begin{align}\label{gammalarges}
&\nonumber\gamma_{s\gg 1}(t)=\int_{-\infty}^{+\infty}\int_{-\infty}^{+\infty}\int_{-\infty}^{+\infty}\int_{-\infty}^{+\infty}dvdwdxdy\,\\
\nonumber&~~~~~~\exp\Big(-\pi(v^2+w^2+x^2+y^2)-2\pi st\big(q_A\noise_{AA}(v-w)^2+2\sqrt{q_Aq_B}\snoise_{AB}(v-w)(x-y)+q_B\noise_{BB}(x-y)^2\big)\Big)\\
&~~~~~~=\frac{1}{\sqrt{1+4st(q_A\noise_{AA}+q_B\noise_{BB})+16s^2t^2q_Aq_B(\noise_{AA}\noise_{BB}-\snoise_{AB}^2)}}.
\end{align}
\end{widetext}
For $s_A=s_B=1/2$ the purity is 
\begin{eqnarray}
&&\nonumber\gamma_{1/2}(t)=(1-q_A)(1-q_B)\\
&&~~~~\nonumber+q_Aq_B\cosh(2\snoise_{AB}t)e^{-\noise_{AA}t-\noise_{BB}t}\\
&&~~~~+q_A(1-q_B)e^{-\noise_{AA}t}+q_B(1-q_A)e^{-\noise_{BB}t}.\label{gammas1/2}
\end{eqnarray} 
The goal of the next section is to compute the optimal purities. 

\section{Purity optimization}

The purity satisfies $\gamma\leq 1$, the inequality being saturated in the absence of decoherence.  
In general the purity cannot be maximized to unity because the condition of positivity of the matrix $C$ in Eq. (\ref{lind}) implies that the noises cannot all vanish simultaneously when the interaction between the subsystems is non-reciprocal. For the setup under consideration, this translates to the quantum noise inequality\cite{Clerk2010}
\begin{equation}\label{ineq}
\noise_{AA}\noise_{BB}\geq\snoise_{AB}^2+\frac{1}{4}\susc_{BA}^2.
\end{equation}
In this section we optimize the purity for the exemplary C-NOT gate for which the time of operation $t$ is given by Eq. (\ref{time}). The optimization procedure is as follows. For a given set of noises $\noise_{AA}$, $\noise_{BB}$, $\snoise_{AB}$, we find the worst initial state that gives minimum purity after the C-NOT operation. This purity depends on the noises and we maximize it over the noise settings subject to the constraint Eq. (\ref{ineq}). This gives us a lower bound for purity that can be achieved by the ideal design. This measure is universal and does not depend on the design details. 

\begin{figure}[h!]
   \includegraphics[width=8cm]{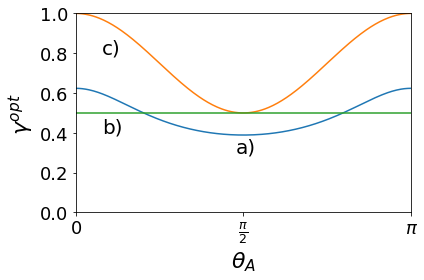}
   \caption{The purities of the C-NOT operation corresponding to three optimal designs in dependence of the initial state. The rotating spin is initially perpendicular to the z-axis ($\theta_B=\pi/2$), $\theta_A$ is the angle between the z-axis and the control spin. Parameters for curve a): $s_A=s_B=s\gg1$, $\noise_{AA}=\noise_{BB}=\susc_{BA}/2$; b): $s_A=s_B=1/2$, $\noise_{AA}\ll\susc_{BA}\ll\noise_{BB}$; c): $s_A=s_B=1/2$, $\noise_{BB}\ll\susc_{BA}\ll\noise_{AA}$.}
\label{figpurities}
\end{figure}

Let us start with the case $s\gg1$. In Eq. (\ref{gammalarges}), all the coefficients in front of $q_A$, $q_B$ and $q_Aq_B$ are positive. This implies that the initial state that yields the worst purity corresponds to maximum $q_A=q_B=1/2$, or $\theta_A=\theta_B=\pi/2$. This purity reads
\begin{eqnarray}
\nonumber\gamma_{s\gg 1}^{min}&=&\big(1+\frac{\pi}{\susc_{BA}}(\noise_{AA}+\noise_{BB})\\
&&+\frac{\pi^2}{\susc_{BA}^2}(\noise_{AA}\noise_{BB}-\snoise_{AB}^2)\big)^{-1/2}.
\end{eqnarray}
It reaches optimum for the choice of the noises $\noise_{AA}^{opt}=\noise_{BB}^{opt}=\susc_{BA}/2$ and $\snoise_{AB}^{opt}=0$, and it equals 
\begin{equation}
\gamma_{s\gg 1}^{opt}=\frac{1}{1+\pi/2}\approx0.39.
\end{equation}
As discussed, this is the bound for the worst initial state. For the same design, the purity for other initial states characterized by $q_A,q_B$ is better: (Fig. \ref{figpurities}) 
\begin{equation}\label{gammaoptlarges}
\gamma_{s\gg 1}^{opt}=\frac{1}{\sqrt{1+\pi(q_A+q_B)+\pi^2q_Aq_B}}.
\end{equation}

The optimization for the case $s_A=s_B=1/2$ is more involved and it is presented in Appendix \ref{g1/2opt}. 
The same optimal purity
\begin{equation}
 \gamma_{1/2}^{opt}=1/2
\end{equation} 
is achieved for two alternative designs. For both designs the initial state giving the worst purity is the same as for the previous case. Its optimum is achieved for $\noise_{AA}\ll\susc_{BA}\ll\noise_{BB}$ or for $\noise_{BB}\ll\susc_{BA}\ll\noise_{AA}$. Although these designs might seem challenging to realize, the purity in this case does not depend much on the precise values of the noises, $\gamma_{1/2}>0.36$ if the noises satisfy $\noise_{AA}\noise_{BB}=\susc_{BA}^2/4$ the limit of the constraint (\ref{ineq}), $\gamma_{1/2}>1/4$ for any noise setting.

For these two designs, we obtain the dependence of the purity on the initial state:

\begin{subequations}
\begin{align}
&\gamma_{1/2}^{opt}=\frac{1}{2}(1+\cos(\theta_A)^2)\text{ for }\noise_{BB}\ll\susc_{BA}\ll\noise_{AA},\label{gammaoptspinhalf1}\\
&=\frac{1}{2}(1+\cos(\theta_B)^2)\text{ for }\noise_{AA}\ll\susc_{BA}\ll\noise_{BB}.\label{gammaoptspinhalf2}
\end{align}
\end{subequations}
In Fig. \ref{figpurities} we plot the purities for optimal designs for $\theta_B=\pi/2$ versus $\theta_A$.

\section{conclusion}

We have studied the interaction between two quantum systems A and B that is mediated by their common linear environment. When the environment is out of thermal equilibrium the interaction in general violates the Onsager symmetries and the action of system A on system B may not produce a back-action. We have derived the corresponding quantum equations and obtained their classical limit. In the quantum domain interaction between the systems without back-action necessarily involves minimal noise on the systems. As an application of the formalism we analyzed the quantum manipulation of spin coherent states of arbitrary spin quantum number and the realization of an analogue of the C-NOT gate. As a measure of the decoherence induced by the noise we evaluated the purity of the quantum states after a time of interaction between the spins that corresponds to the time of operation of the C-NOT gate. The final purity depends on the initial states of the spins and we optimized the purity for all initial states. In the worst case of the initial state the optimal purity is $1/2$ for spin 1/2 and $\approx 0.39$ for large spins. Thus even though the decoherence is important for two-level systems, it can be made relatively small for systems of large spins. Similar considerations can be applied to more complex quantum gates and systems. The non-symmetric Onsager interaction can be used to perform quantum operations. However it always brings decoherence. The minimal decoherence depends on the initial state. The worst minimal decoherence may become of the order 1 for simple two-level systems. However the decoherence per quantum degree of freedom can be significantly reduced by employing larger quantum systems as we have illustrated with the example of large spins.

\emph{Acknowledgements.} 
We acknowledge interesting discussions with Lieven Vandersypen. This work was partly supported by the European Research Council (ERC-Synergy).

\appendix

\section{Purity optimization for $s=1/2$}\label{g1/2opt}
Here we derive the optimal purity for $s=1/2$. Eq. (\ref{gammas1/2}) is of the form
\begin{equation}\label{gamma1/2}
\gamma_{1/2}(t)=1-aq_A-bq_B+cq_Aq_B,
\end{equation}
with
\begin{subequations}\label{def}
\begin{align}
&a=1-\exp(-t\noise_{AA}),\\
&b=1-\exp(-t\noise_{BB}),\\
\nonumber&c=1+\cosh(2 t \snoise_{AB})\exp(-t\noise_{AA}-t\noise_{BB})\\
&-\exp(-tN_{AA})-\exp(-t\noise_{BB}).
\end{align}
\end{subequations}

Eq. (\ref{gamma1/2}) defines hyperbolas in the plane $(q_A,q_B)$ whose center is at $(q_A,q_B)=(b/c,a/c)$. Based on this observation it is straightforward to find the minima of the purity (\ref{gamma1/2}) in the domain $q_A,q_B\in[0,1/2]$, depending on the parameters $a,b\text{ and }c$. We summarize the expressions of the minimal purities and their locations in the table:
\begin{center}
\begin{tabular}{ c  c  c }
  \hline
  \hline
  \noalign{\smallskip}
  $\gamma_{1/2}^{min}(t)$ & Conditions & $(q_A,q_B)$\\
  \noalign{\smallskip}
  \hline
  \noalign{\smallskip}
  $1-\frac{a}{2}-\frac{b}{2}+\frac{c}{4}$ & $a\geq\frac{c}{2}$, $b\geq\frac{c}{2}$ & $(\frac{1}{2},\frac{1}{2})$\\
  $1-\frac{a}{2}$ & $b\leq\frac{c}{2}$, $a\geq b$ & $(\frac{1}{2},0)$\\
  $1-\frac{b}{2}$ & $a\leq\frac{c}{2}$, $b\geq a$ & $(0,\frac{1}{2})$\\
  \noalign{\smallskip}
  \hline
  \hline  
\end{tabular}
\end{center}

Then we maximize the purity with respect to the noises at time $t_{\pi/2}$. We have done this numerically using the scipy.optimize package\cite{scipyoptimize} and we found that the optimal purity is obtained when either $\noise_{AA}\to 0$ and $\noise_{BB}\to\infty$ or $\noise_{AA}\to\infty$ and $\noise_{BB}\to0$, and it is equal to $1/2$. 

\begin{figure}[h!]
   \includegraphics[width=8cm]{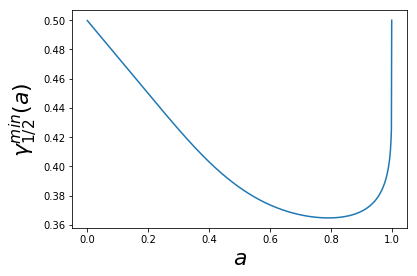}
   \caption{Purity Eq. (\ref{ga}) as a function of $a$, Eq. (\ref{def}), and for $t=t_{\pi/2}$.}
\label{figga}
\end{figure}

In order to see this we let $\snoise_{AB}=0$. This gives $c=ab$ and it corresponds to the conditions $a\geq\frac{c}{2}$ and $b\geq\frac{c}{2}$. Then the minimal purity is obtained for $q_A=q_B=1/2$ and it is equal to
\begin{equation}\label{gmin}
\gamma_{1/2}^{min}(t)=\Big(1-\frac{a}{2}\Big)\Big(1-\frac{b}{2}\Big).
\end{equation}
The noises that maximize the purity (\ref{gmin}) saturate the quantum noise inequality (\ref{ineq}). At time $t_{\pi/2}$ (\ref{time}) they are related by
\begin{equation}
\ln(1-a)\ln(1-b)=\frac{\pi^2}{4}.
\end{equation}
With this we express $b$ as a function of $a$, insert the expression into Eq. (\ref{gmin}), and we obtain
\begin{equation}\label{ga}
\gamma_{1/2}^{min}(a)=\frac{1}{2}\Big(1-\frac{a}{2}\Big)\Big[1+\exp\Big(\frac{\pi^2}{4\ln(1-a)}\Big)\Big].
\end{equation}
We plot Eq. (\ref{ga}) in Fig. (\ref{figga}).
The maxima of Eq. (\ref{ga}) are obtained for $a=0\text{ and }1$, that is when the noise $N_{AA}$ either vanishes or diverges, and the optimal value of the purity is $1/2$.


\begin{thebibliography}{9}

\bibitem{Wolf2001} S. A. Wolf, D. D. Awschalom, R. A. Buhrman, J. M. Daughton, S. von Molnar, M. L. Roukes, A. Y. Chtchelkanova, and D. M. Treger, Science {\bf 294}, 1488 (2001). 

\bibitem{Awschalom2002} D. Awschalom, D. Loss, and N. Samarth (ed.), Semiconductor Spintronics and Quantum Computation, Springer (2002). 

\bibitem{Zutic2004} I. Zutic, J. Fabian, and S. Das Sarma, Rev. Mod. Phys. {\bf 76}, 323 (2004).

\bibitem{QuantumInformation} M. A. Nielsen and I. L. Chuang, \textit{Quantum Computation and Quantum Information}, Cambridge University Press, 10th Anniversary Edition, 2010.

\bibitem{LDV} D. Loss and D. P. DiVincenzo, Phys. Rev. A {\bf 57}, 120 (1998).

\bibitem{Kane} B. E. Kane, Nature {\bf 393}, 133 (1998).

\bibitem{HansonRMP} R. Hanson, L. P Kouwenhoven, J. R. Petta, S. Tarucha, and L. M. K. Vandersypen, Rev. Mod. Phys.
{\bf 79}, 1217 (2007).

\bibitem{ZwanenburgRMP} F. A. Zwanenburg, A. S. Dzurak, A. Morello, M. Y. Simmons, L. C. L. Hollenberg, G. Klimeck, S. Rogge, S. N. Coppersmith, and M. A. Eriksson, Rev. Mod. Phys. {\bf 85}, 961 (2013).

\bibitem{Moiseyev} N. Moiseyev, Non-Hermitian Quantum Mechanics, Cambridge University Press, Cambridge,
(2011).

\bibitem{Uzdin2013} R. Uzdin, J. Phys. A: Math. Theor. {\bf 46}, 145302 (2013).

\bibitem{Muller2012} M. M\"uller, S. Diehl, G. Pupillo, P. Zoller, Advances in Atomic, Molecular, and Optical Physics {\bf 61}, 1 (2012). 

\bibitem{Emary2013} C. Emary, Phys. Rev. A {\bf 87}, 032106 (2013).

\bibitem{Bender2010} C. M. Bender, D. C. Brody, J. Caldeira, and B. K. Meister, arXiv:1011.1871.

\bibitem{Evans2015} D. E. Evans and T. Gannon arXiv:1506.03546.

\bibitem{Ketterer2014} A. Ketterer, S. P. Walborn, A. Keller, T. Coudreau, and P. Milman, arXiv:1406.6388.

\bibitem{Petta2005} J. R. Petta, A. C. Johnson, J. M. Taylor, E. A. Laird, A. Yacoby, M. D. Lukin, C. M. Marcus, M. P.
Hanson, and A. C. Gossard, Science {\bf 309}, 2180 (2005).

\bibitem{Niskanen2007} A. O. Niskanen, K. Harrabi, F. Yoshihara, Y. Nakamura, S. Lloyd, and J. S. Tsai, Science {\bf 316},
723 (2007).

\bibitem{Nowack2011} K. C. Nowack, M. Shafiei, M. Laforest, G. E. D. K. Prawiroatmodjo, L. R. Schreiber, C. Reichl, W. Wegscheider, and L. M. K. Vandersypen, Science
{\bf 333}, 1269 (2011).

\bibitem{Braakman2013} F. R. Braakman, P. Barthelemy, C. Reichl, W. Wegscheider, L. M. K. Vandersypen, Nature Nanotechnology {\bf 8}, 432-437 (2013)

\bibitem{Srinivasa2015} V. Srinivasa, H. Xu, and J. M. Taylor, Phys. Rev. Lett. {\bf 114}, 226803 (2015).

\bibitem{Baart2016} T. A. Baart, T. Fujita, C. Reichl, W. Wegscheider, and L. M. K.
Vandersypen, Nat. Nano. {\bf 11}, 330 (2016).

\bibitem{Kubo} R. Kubo, J. Phys. Soc. Jpn. {\bf 12}, 570 (1957). 

\bibitem{Onsager} L. Onsager, Phys. Rev. {\bf 37}, 405 (1931).

\bibitem{VanKampen} N. G. Van Kampen, Stochastic Processes in Physics and Chemistry, Amsterdam, Elsevier, (2011).

\bibitem{Benatti2003} F. Benatti, R. Floreanini, and M. Piani, Phys. Rev. Lett. {\bf 91}, 070402 (2003).

\bibitem{Safi} I. Safi and P. Joyez, Phys. Rev. B {\bf 84}, 205129 (2011) and I. Safi in "Noise and Fluctuations", AIP Conference Proceedings of XX Int. Conf. on Noise and Fluctuations (Pisa), Vol. 1129, edited by M. Macucci and G. Basso (AIP, Melville, NY 2009).

\bibitem{Lai2008} C-Y. Lai, J-T. Hung, C-Y. Mou, and P. Chen, Phys. Rev. B {\bf 77}, 205419 (2008).

\bibitem{Sun2018} Z. Sun, X-Q. Xu, and B. Liu, Phys. Rev. A {\bf 97}, 052309 (2018).

\bibitem{Monroe1995} C. Monroe, D. M. Meekhof, B. E. King, W. M. Itano, and D. J. Wineland, Phys. Rev. Lett. {\bf 75}, 4714 (1995).

\bibitem{Veldhorst2015} M. Veldhorst,  C. H. Yang, J. C. C. Hwang, W. Huang, J. P. Dehollain, J. T. Muhonen, S. Simmons, A. Laucht, F. E. Hudson, K. M. Itoh, A. Morello, and A. S. Dzurak, Nature {\bf 526} 410 (2015).

\bibitem{Zajac2018} D.M. Zajac, A. J. Sigillito, M. Russ, F. Borjans, J. M. Taylor, G. Burkard, J. R. Petta, Science {\bf 359}, 439 (2018).

\bibitem{Danon2011} J. Danon and Y. V. Nazarov, Phys. Rev. B {\bf 83}, 245306 (2011).

\bibitem{Lieb73} E. H. Lieb, Commun. math. Phys. {\bf 31}, 327 (1973).

\bibitem{Clerk2010} A. A. Clerk, M. H. Devoret, S. M. Girvin, F. Marquardt, and R. J. Schoelkopf, Rev. Mod. Phys. {\bf 82}, 1155 (2010).

\bibitem{scipyoptimize} https://docs.scipy.org/doc/scipy/reference/generated/ scipy.optimize.minimize.html

\end{thebibliography}
\end{document}